\documentclass[%
 reprint,
superscriptaddress,
longbibliography,
showpacs,
nofootinbib,
 amsmath,amssymb,aps,
prd,
]{revtex4-2}

\nocite{TitlesOn}
\usepackage{lipsum} 

\newcommand{\cmmnt}[1]{}
\usepackage{bm}
\usepackage{physics}
\usepackage{comment}
\usepackage{floatrow} 
\pdfoutput=1
\usepackage{amsmath,amsfonts,amsthm}
\usepackage{esdiff}  
\usepackage{booktabs}  
\usepackage{url}  
\usepackage{hyperref}  
\usepackage{relsize}
\usepackage{cleveref}  
	\crefname{equation}{equation}{equations}
	\crefname{figure}{figure}{figures}	
	\crefname{table}{table}{tables}
\usepackage[caption=false]{subfig}

\usepackage[normalem]{ulem}

\usepackage[usenames,dvipsnames,svgnames,table]{xcolor}

\usepackage{graphicx}

\usepackage{aurical}
\usepackage[T1]{fontenc}

\usepackage[sc]{mathpazo} 
\usepackage[T1]{fontenc} 
\usepackage{microtype} 

\usepackage{booktabs} 
\usepackage{float} 
\usepackage{hyperref} 

\usepackage{lettrine} 
\usepackage{paralist} 

\definecolor{darkred2}{HTML}{880808}
\definecolor{crimson}{HTML}{DC143C}

\usepackage{titlesec} 
\renewcommand\thesection{\Roman{section}} 
\renewcommand\thesubsection{\Alph{subsection}} 
\titleformat{\section}[block]{\large\scshape\centering\bfseries}{\thesection.}{1em}{} 

\titleformat{\subsection}[block]{\scshape\centering}{\thesubsection.}{1em}{} 

\DeclareCaptionFormat{myformat}{#1#2#3\hrulefill}
\usepackage{float}
\floatstyle{plaintop}
\restylefloat{table}

\begin{document}
\nocite{TitlesOn}

\title{What's Chasing Me?}
\title{Why Am I Running When There's Nothing Chasing Me?}
\title{What Am I Running From?}
\title{What Am I So Mad About?}
\title{Why Am I So Mad When Nothing's Wrong?}
\title{Why Do Things Upset Me For No Reason?}
\title{What Am I So Irritated About?}
\title{Why Am I So Edgy For No Reason?}
\title{Why Am I So Mad When Nothing Happened?}
\title{Why Am I So Edgy When Nothing's the Matter?}
\title{I'm okay.}
\title{Everything's Fine.}

\author{Eve Armstrong\thanks{earmstrong@amnh.org}}
\email{earmstrong@amnh.org}
\affiliation{Department of Physics, New York Institute of Technology, New York, NY 10023, USA}
\affiliation{Department of Astrophysics, American Museum of Natural History, New York, NY 10024, USA}
\affiliation{\url{http://www.amazon.com/author/evearmstrong}}

\date{April 1, 2024}

\begin{abstract}
I investigate the peculiar situation in which I find myself healthy and strong, with a darling family, stimulating job, top-notch dental plan, and living far from active war and wildfire zones -- yet perpetually ill at ease and prone to sudden-onset exasperation when absolutely nothing has happened.  My triggers include dinner parties, chairs, therapists, and shopping at Costco.  In analysing this phenomenon, I consider epigenetics, the neuroscience of neuroticism, and possible environmental factors such as NSF grant budgets.  Yet no obvious solution emerges.  Fortunately, my affliction isn't really all that serious.  In fact, it's good writing material.  So while I'm open to better ideas, I figure I'll just continue being like this. 
\end{abstract}

\maketitle

\section{Introduction} \label{sec:intro} 

Although I hide it well, I have a tendency to become anxious and irritable under what most would consider benign circumstances.  Edgy, unnerved, even a little spooked.  Sometimes downright enraged.  For no clearly identifiable reason.

\begin{figure*}
\includegraphics[width=0.8\textwidth]{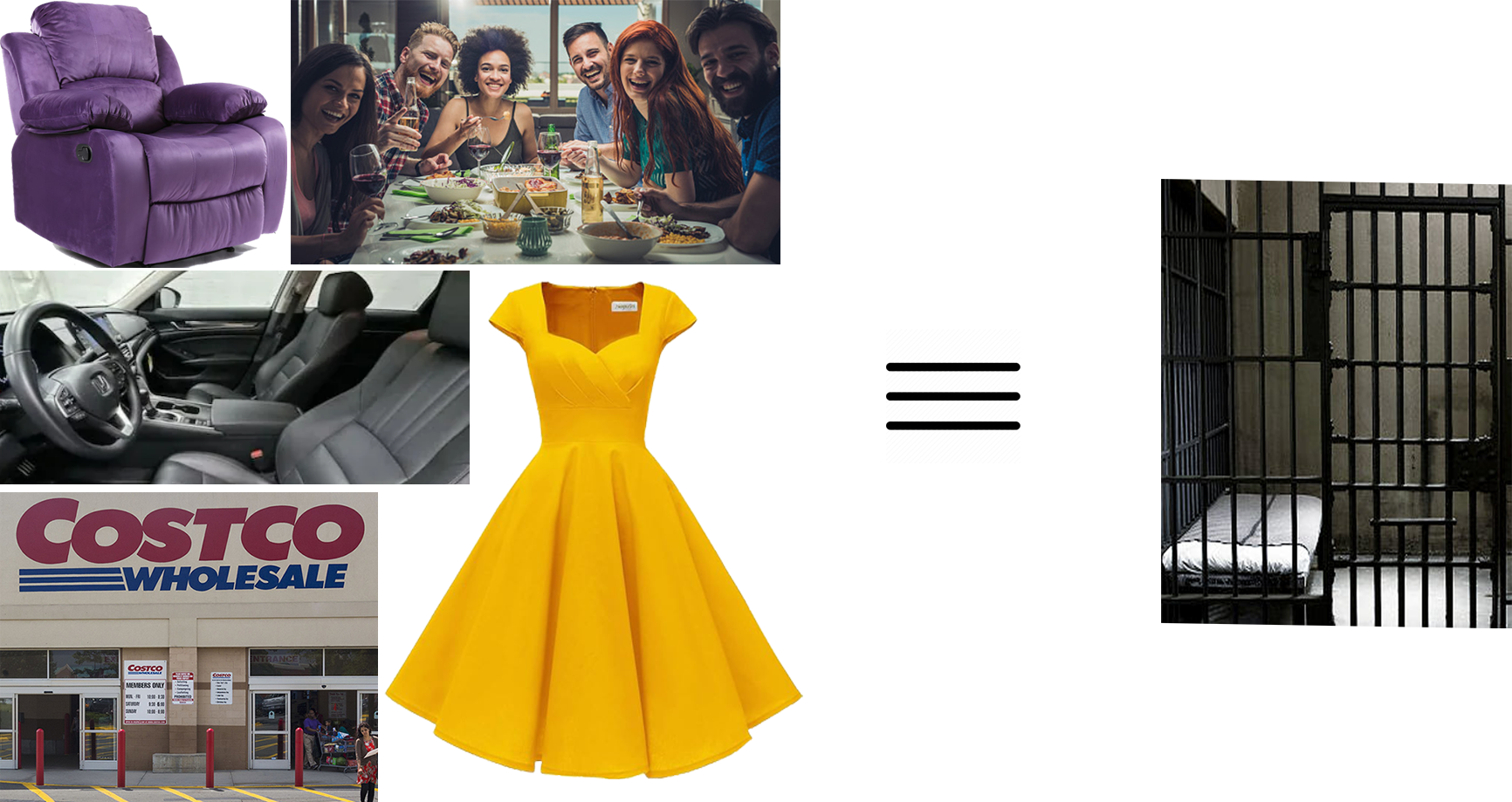}
\caption{\textit{Left}: \textbf{Some of my triggers.}  \textit{Clockwise from top left:} sitting in a low cushioned chair~\cite{chair}, dinner parties~\cite{dinnerParty}, wearing a dress~\cite{dress}, shopping at Costco~\cite{Costco}, riding in a car~\cite{car}.  \textit{Right:} \textbf{What these triggers represent to me~\cite{prison}.}} 
\label{fig1}
\end{figure*}

Sitting still is particularly upsetting\footnote{Thankfully, I sleep like the dead.  So long as I'm unconscious, I'm fine with it.}.  The worst are low cushioned seats that suck you in like quicksand.  And it's not the stillness alone -- it's the social expectation that I will be still.  At a dinner party, for example.  There, in addition to sitting still, I am expected to be happy about it for an extended duration --- \textit{and} simultaneously to care that the eight-year-old daughter of this woman Jill I've just met is in the Gifted and Talented Program at Wolcott Elementary\footnote{Also that Jill is into Taekwondo and it's helped with her assertiveness in the workplace, and also that she and her husband did Ayahuasca while they were in Mexico, and it was amazing and I have to try it.  Meanwhile, I feel even more burdened that Jill has to pretend to care about my upcoming conference on neutrino physics in Florence.}.  Other constraints that are hard to deal with include being in a car, spaces with no window or exit in sight -- like the Costco wholesale warehouse~\cite{Costco}, and wearing a dress (there's a "prison" motif here; see Fig.~\ref{fig1}).  Essentially, I panic whenever I look around and realize that taking off in a sprint would be tricky.  

I also struggle to focus for long in a passive manner.  A movie, even one I find compelling, I can only inch through in 25-minute installments.  Then busy weekdays intervene and when there's finally time to continue, I have to rewind ten minutes to remind myself.  Wayne, my poor husband: he loves old classics.  We started \textit{Citizen Kane} three weeks ago and we're only an hour in.  

Other events incite, not anxiety so much, but irritation.  Wayne and I arrive at the subway platform to have the doors close in our faces, and I'm annoyed.  Even though it's Saturday and we're just heading for a grocery run at Trader Joe's.  Then Wayne grins at the awkward timing, and my annoyance flares to anger.  I have to check myself from snapping at him, "What are you so happy about?"  Even in the moment, I retain the rationality to think: \textit{Eve, what is wrong with you?  Are you really angry that Wayne isn't angry?}  Probably not.  Probably it's something else.

But what?

\begin{figure}
\includegraphics[width=0.5\textwidth]{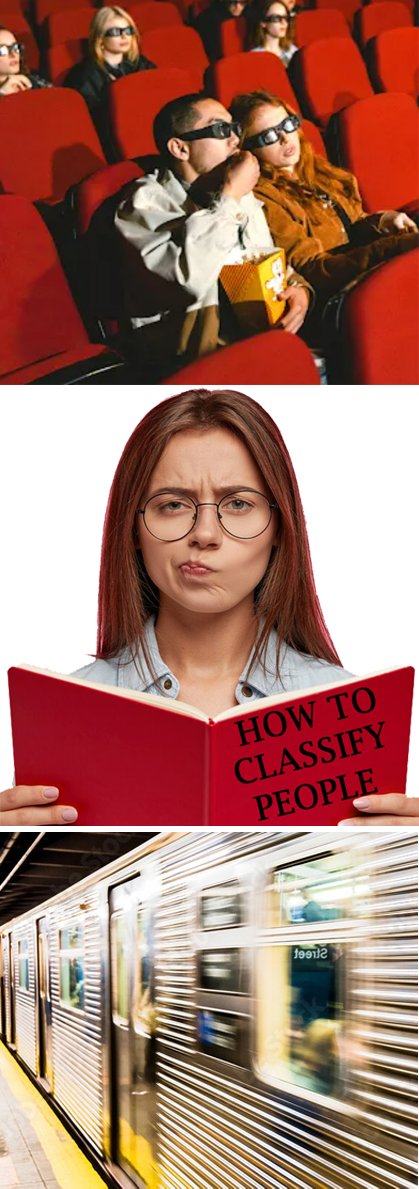}
\caption{\textbf{Other triggers.}  \textit{From top:} watching an entire movie in one sitting~\cite{movies}, therapists~\cite{therapist}, missing the subway~\cite{subway} even if I have nowhere important to go.} 
\label{fig2}
\end{figure}

Talk therapy is an evidence-based intervention to understand one's triggers and desensitize oneself to them~\cite{wenzel2017basic}.  Many people find talk therapy helpful (e.g. Ref.~\cite{hofmann2012efficacy}), and that is wonderful for them.  Whatever solution works.  For me, therapists are yet another trigger (for some of my non-prison-motif triggers, see Fig.~\ref{fig2}).  Talking about my problems heightens them.  And I'm not easy about the way therapists have been trained to classify me, nor what their motivations might be in doing so.

Things I do find helpful: running.  Camouflaging in fiction; say, writing a character to do a version of the thing that would make trouble were I to do it myself.  Writing privately about how ridiculous I am in being like this.  Noting ridiculous things happening all around me, and that I'm in good company.  In addition, talk therapy has it right about avoiding triggers.  Given the typical-everyday nature of mine, however, avoidance is often impractical.  Alas, would that my triggers were serial killers and Sweden\footnote{That's not a dig at Sweden; I just never go there.} rather than chairs and people.  

So can I deactivate these triggers, or at least water them down?  Maybe -- there exists a wealth of literature that might help.  Possibly-relevant disciplines include psychology, neuroscience, and epigenetics.  These fields differ widely in methodology, yet they dig at the same question: what makes us us -- and can we do anything about it?   

\section{Possibility 1: It's environmental.} \label{sec:theory1}

\subsection*{\textbf{Prior trauma?}}

Students are reportedly stressed out.  PhD students are statistically more likely to report mental-health-related issues than the general population~\cite{friedrich2023your,evans2018evidence}.  As a PhD student, I never reported mental health issues to anybody, but it's possible that I had some.

I went to graduate school after two years as an actor/writer in New York City (following my undergrad years) taught me that I'd like to have health insurance.  Physics was the obvious Plan B.  I savored the feel the problem sets gave, warping my brain's prior configuration.  It hurt good.  Like challenging a muscle you didn't know you had.  And it would pay.

But the brain-warping did not come naturally for me.  During my first trimester at the University of California, San Diego, I kept Sakurai's \textit{Modern Quantum Mechanics} at my bedside along with a notepad and pencil, so that -- when I woke at 2am in a panicked sweat -- I wouldn't miss a beat.  Meanwhile, the southern-California culture mystified me.  No one seemed to appreciate sarcasm.  So I stopped using it, which was deeply disquieting, as it has been my family's primary form of communication for generations\footnote{In fact, the atrophy of my sarcasm synapses probably altered my epigenome; see Section~\ref{sec:theory3}.}.   

Then came postdoc years, still in San Diego.  Again, fascinating work but stressful lifestyle.  Wayne was maintaining our NYC apartment, so to avoid rent, I lived in my campus office.  I slept on two dog beds from Petco.  All the while I wondered, \textit{What comes after this?  If I can't land a faculty job, what will my platform be for lampooning academia on April Fool's Day?}  

\subsection*{\textbf{Current trauma?}}

I struck gold with a faculty job and affiliation with the broader NYC astro/physics community.  The science, the platform to do comedy, an eclectic bunch of students and colleagues -- it's all here.  Still, perhaps there exist sources of stress.  The literature on stress in faculty (e.g. Refs~\cite{higgins2006factors,ashrafi2014study}) is much sparser than that on students.  In particular, I have found no studies on the psychological effects of spending 20\% of one's work week on tedium.    

Let's take meetings.  During the COVID-19 pandemic lockdown, the advent of Zoom~\cite{zoom} -- a video-telephony software program -- significantly enhanced meeting popularity.  It is a double-edged sword that that popularity has stuck.  Some meetings are important, and Zoom offers ease and flexibility.  On the other hand, as I type this, I am twelve minutes into a Zoom meeting wherein so far we have discussed: 1) whether anyone would like to suggest a meeting agenda, and 2) when our next meeting shall be.

Grant upkeep also eats nerves.  I have funding from the National Science Foundation (NSF): it is a competitive grant, and I am honored to be trusted that I will do right by the scientific community.  And with that trust comes responsibility.

There is an expectation to publish regularly.  Now, to be fair, I can't think of a handier metric to keep tabs on all the theoretical-physicist grantees the NSF must monitor.  Our job is to think.  Besides production rate, how do you measure how hard a person has thought over the past year? 

Well, despite thinking as hard as we usually do, progress this year has been slow for our research group.  We don't quite have a publication-ready result.  That's been stressful.  Luckily, we've recently realized that while the NSF will note whether we have a new publication, they won't actually read it.  So we now have a paper in preparation, tentatively titled: "Pretty Much the Same Result We Published Last Year" (in preparation for submission to the \textit{Physical Review D}).

Keeping this grant also requires \textit{using} the grant, which is a monster all on its own.  The grant goes not to you the grantee, but to your institution.  Then you must convince the institution to let you have some of it.

For example, my grant supports travel to visit scientific collaborators.  Once, using a credit card linked to the grant, I bought a flight protection policy with Delta Airlines.  I bought it because on my previous trip, the return segment had been canceled due to weather and the airline hadn't re-booked me, which cost my grant half the $\sim$ \$500 ticket.  So, in purchasing this \$19.99 policy, I congratulated myself for learning my lesson and acting responsibly.  But nope: my institution does not permit the use of external grants for flight protection.  After an hour's argument, I calculated how much of my salary I had just used up in claiming superior logic, and wrote a personal check for \$19.99.

As my research is in computational physics, my grant also covers a computer.  I prefer a Hewlett-Packard (HP) laptop: I have a good history with the brand, and they're light and thin enough to bring anywhere.  So I bought one.  But this was a crime -- because my institution has a business agreement with Dell, a competitor of HP\footnote{To those who may be angry that I'm putting this in writing: hey, either I at least get a story out of it, or I'll be slowly growing a tumor somewhere.}.  I spent the equivalent of two work days defending my decision to use my grant to buy the thing I use to do the work the grant pays for.  I wound up narrowly avoiding having to pull out my checkbook again, this time for \$1300\footnote{I also narrowly avoided commenting: "It is not the NSF's fault that you're sleeping with Dell."}. 

With a grant, each expense is suspect.  Eventually, on some level you start believing it.  Making off with a \$6 bag of peanuts from La Guardia Airport feels criminal, even after you spend an honest ten minutes chasing down the receipt for the reconciliation form.  

But I shouldn't complain.  Of course the university requires accountability over access to a pile of cash.  And how grateful I am for what I get to do as a result of the toil!  To offer enthused students the kinds of opportunities I was lucky enough to have, to work alongside talented colleagues, to spend all of July at a scientific conference in Florence -- it's all worth it.

... Right?

\subsection*{\textbf{This doesn't add up.}}

Well, mulling all this over, I'm not buying that my problem is simply environmental.  The notion that academia is inordinately stressful is ridiculous.  What about in coal mining?  You don't hear that making the news.  The New York Times has never published a piece on job-related anxiety in coal miners\footnote{Nor on loggers, cattle thieves (see Section~\ref{sec:theory3}), or derrick operators.}.  The literature cited in this Section is based largely on self-reported feelings and observed behavior.  Subjective.  Do academics really feel more anxious than the general population -- or are we more elitist, entitled, and noisier than the general population\footnote{And/or are we not more likely to have both the health insurance and the time to complain to a therapist?  And/or does academia tend to attract people disposed to complaining to therapists?}?

I complain, but in truth I am astounded to have this job\footnote{Also, I've been a Manhattan real estate agent and a TJ Maxx salesperson specializing in ladies' underwear -- and was no more or less anxious then than now.}.  I get a month working in Italy this summer on the government's tab!  Shouldn't I expect some overhead?!  Anyway, if I am indeed traumatized by tedium, why would it manifest as an aversion to chairs?

\section{Possibility 2: It's me.}\label{sec:theory2}

If my environment is not the problem, maybe it's just me.  By that I do not mean imposter syndrome: I am perfectly fine with being an imposter right along with all of you.  And I consider myself to be in excellent company. 

By "me" I mean my intrinsic wiring.  Personality neuroscience~\cite{deyoung2022personality} is relevant here: it's a sub-discipline of neuroscience that employs molecular genetics, pharmacological assays, electroencephalography (EEG), and neuroimaging to identify neurobiological mechanisms underlying individual personality differences.  

Many researchers describe personality in terms of five basic traits~\cite{wiggins1996five}: agreeableness, conscientiousness, openness to experience, extroversion, and neuroticism.  Let's home in on neuroticism: the tendency to experience negative emotions, including anxiety and irritability.  It is a risk factor for psychopathology~\cite{deyoung2022personality}.  In other words, those who score high on neuroticism are more likely than average to become burglars or U.S. presidents.  

In human EEG studies, neuroticism has been tied to aberrant electrical activity in the brain area called the amygdala~\cite{fox2019central}.  A correlation has also been found between neurotic tendencies and enhanced activity in the dorsal anterior cingulate cortex~\cite{saunders2020assessing}.  In large human samples, a negative correlation has been found between neuroticism and the surface area of the dorsomedial prefrontal cortex, a region that might be involved in the subjective experience of emotion~\cite{grasby2020genetic} (there are a lot of "mights" in this literature).

Those references on human brain studies seem a bit cagey on details.  But to be fair, to dig in deep is illegal.  One study in mice was more specific: heightened anxiety was claimed to be caused by enhanced excitatory synaptic inputs onto somatostatin-expressing neurons in the central amygdala, and a resulting reduction in inhibition onto the bed nucleus of the stria terminalis~\cite{ahrens2018central}.  The relevance to human emotion, however, is unclear.

\subsection*{\textbf{This doesn't help either.}}

Not to denigrate the science, but I don't see where this is getting me.  I do not believe I abhor chairs due to insufficient dorsomedial prefrontal cortex surface area.  Compared to correlation, causation is fiercely difficult to ferret out.  Especially if the system is nonlinear, as the central nervous system certainly is.

\section{Possibility 3: It's both.} \label{sec:theory3}

Speaking of nonlinearity, let's look at epigenetics.  This I find fascinating: that there exists biological evidence to corroborate my sense (and possibly yours?) that the relationship between an organism's behavior and environment is inherently nonlinear.  Or: your behavior and environment can affect your behavior in that environment.

Every biological cell possesses DNA, a hard-coded sequence of genes.  But there's more to it; otherwise identical twins should have indistinguishable personalities.  Epigenetics is part of the "more": the means by which a cell expresses or silences these genes (e.g. Ref~\cite{carey2012epigenetics}).  

DNA interacts with other chemicals within the cell -- the cell's "epigenome" -- that weight each gene's contribution.  A gene that is expressed will generate proteins that determine the cell's function, while a gene that is silenced will not contribute.  One's heart and brain cells have identical DNA but different epigenomes -- and different roles.  Put simply: DNA is the playscript, and epigenetics is the production: the choice of actors, amount of rehearsal time, direction, lighting, audience demographic, choice of snacks at intermission, and whether they cut the clich\'e dream sequence from Act II.

Also unlike DNA, the epigenome can evolve.  Evidence exists that epigenome changes can be caused by environment, including social interactions, and the organism's own behavior~\cite{moore2017behavioral,seebacher2019epigenetics,feil2012epigenetics}.  One well-studied environmental effect is tolerance to alcohol.  Upon repeated alcohol use, one's liver cells crank up the expression of the gene required to break it down~\cite{berkel2017emerging,krishnan2014epigenetic}\footnote{Given the right spin, this is scientific justification for buying twice as much tequila as you usually do.  I hope the liquor companies appreciate the free advertising.}.  Some epigenetic changes can survive cell division, and so potentially last for an organism's life; some can be reversed.  

Let's bring in anxiety.  Evidence for a causal relationship between epigenetics and anxiety comes largely from rodent studies.  In one (Ref.~\cite{weaver2004epigenetic}), rodent babies who were neglected by their mothers displayed heightened stress response to benign stimuli as adults, compared to babies raised by nurturing mothers.  Following nurturing during adolescence, the effect reversed to some degree.    There is also evidence from human studies for epigenetic modulation of stress responses~\cite{puglia2015epigenetic}, although it is controversial.

More intriguing is evidence that parents can pass environmentally-acquired epigenetic changes to their offspring.  In a widely cited study~\cite{dias2014parental}, male mice were exposed to the scent of a chemical reminiscent of cherry blossoms, while given an electric shock.  They learned to associate the scent with the shock, and would tremble upon smelling it even when the shock did not occur.  The mice were then bred.  At birth, their offspring would tremble at that scent, absent a shock.  Upon dissection, those offspring had the same epigenetic modifications their fathers had developed, in the brain cells involved with odor detection.  There exists controversial evidence that such "transgenerational trauma" can extend to the third generation.  Criticisms include small sample sizes and failure to note which offspring were genetically related~\cite{krippner2019transgenerational}.  Despite the controversy, I am intrigued by a possible biological explanation for why my nephew would recoil at a mosquito from the month he was born, even though he'd never seen \textit{Jurassic Park}. 

Is there anything in epigenetics that can explain my own anxiety?  My mother is the opposite of a neglectful rat mother: attentive and doting.  And my father does not recall having been repeatedly shocked by a low cushioned chair prior to having me.  But my parents -- who are lovely and loving -- are also tense.  More tense than me.  It was a tightly wound household growing up.  My dad says his dad was worse.  

\begin{figure}[htb]
\includegraphics[width=0.4\textwidth]{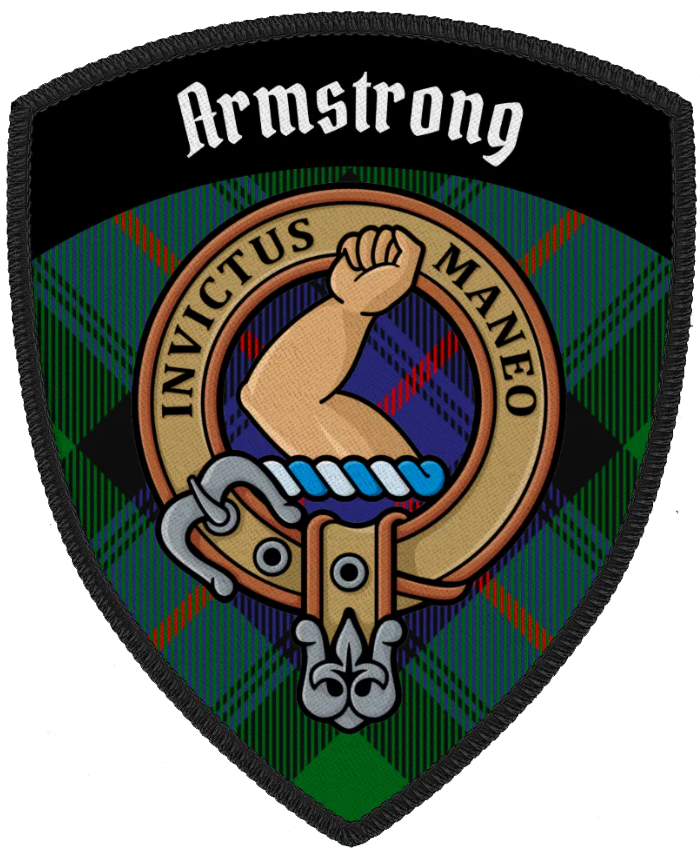}
\caption{\textbf{The Armstrong clan's coat of arms~\cite{coatOfArms}.} \textit{Invictus Maneo} means: "I remain unvanquished."  So it seems that I stem from an irritable lot.  (My dad says our clan branch's version has blood spurting from the arm, but Google didn't turn it up).} 
\label{fig3}
\end{figure}

Further, Dad's grandfather is just one generation out from the "lawless" Armstrong clan of Scotland.  He immigrated to the U.S. at 14, and had been kept ignorant of his close kin's occupation.  Years later on a visit back to his homeland, he spotted a castle and asked a local farmer who lived there.  "Sir William Armstrong," the farmer told him.  "Biggest cutthroat cattle thieves in the countryside."  Literature supports this: Ref.~\cite{rutledge1966lawless} states, "In and around the 'Debateable Land' that lay between the Kingdoms of Scotland and England, the most important clan in both numbers and notoriety was that of Armstrong ... they were uniformly lawless ... and in time came to be labelled officially as freebooters, robbers, murderers, and thieves."  Our family coat of arms substantiates that description (Fig.~\ref{fig3}).  

\subsection*{\textbf{There may be something to this.}}

I speculate that cutting throats and stealing cattle is a high-irritability occupation.  Not only due to the job, but also a job requirement: wouldn't it be hard to be both an effective murderer \textit{and} a calm one?

Further, I think I can feel Sir William Armstrong's blood in me.  Given the proper conditioning, I can imagine myself a murderer.  What an efficient way to dissipate rage, compared to writing a book or running twelve miles.  I'm all for handy time-savers.

Okay, then: a thought experiment.  Say irritability "runs in my family" in that -- on comparing my epigenome to that of my great great granddad -- similarities exist in our brain cells associated with emotion modulation, and that statistically, those similarities are significantly different from the general population.  And let's buy causality: epigenetic transmission.

But what was the mechanism?  Was it transgenerational inheritance, like the mice whose dads were shocked while smelling cherry blossoms?  Or was it environment, by means of having been reared by Sir William Armstrong's great grand-relative?  Wow: it seems the nature-versus-nurture paradigm needs dropping the "versus" (e.g. Ref~\cite{wright2022nature}). 

Even if those two possible routes of transmission were distinct, we can't test for either.  We could examine whether murdered bodies are among my triggers, but that'd be inconclusive: they're probably among most people's triggers.  Nor can we rewind and have me adopted at birth by the Anglican Pacifist Fellowship~\cite{anglican}, and see how I turn out.

Well, mechanism aside, the Armstrong irritability has been watering down.  After their heyday around 1650~\cite{stewart2017armstrongs}, our branch married into the MacKenzie clan, who by all accounts were a civil lot~\cite{mackenzie1879history}.  I, four generations out from there, do not steal chairs or murder people at Costco\footnote{Or at dinner parties (in case you're worried about lies of omission).}.  Given my lineage, I should just be grateful that I'm able to keep my angst to myself. 

... Right?

\setlength{\tabcolsep}{5pt}
\begin{table}[htb]
\small
\centering
\begin{tabular}{p{2.8cm} b{2.2cm} | p{2.5cm}}
\toprule
 \textbf{Possible reason ...} &  & \textbf{... which doesn't add up} \\\midrule \hline
 \raisebox{-.65\height}{\includegraphics[width=0.22\textwidth]{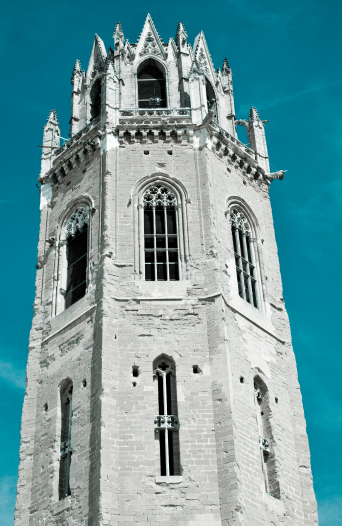}} & \textit{Environment}: Living in the ivory tower (Sec.~\ref{sec:theory1}). & But ivory-tower stress is nothing compared to, say, cattle-thieving stress (I've heard). \\ \hline  
 \includegraphics[width=0.35\textwidth]{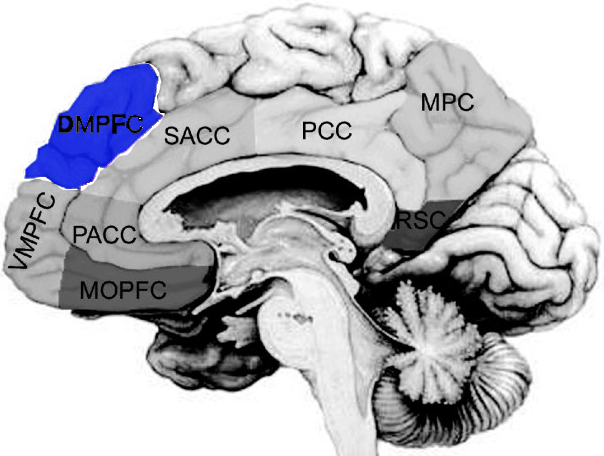} & \textit{Intrinsic wiring}: Insufficient dorsomedial prefrontal cortex surface area - the blue at left (Sec.~\ref{sec:theory2}).  & But it's hard to show causality. \\ \hline
 \raisebox{-.2\height}{\includegraphics[width=0.3\textwidth]{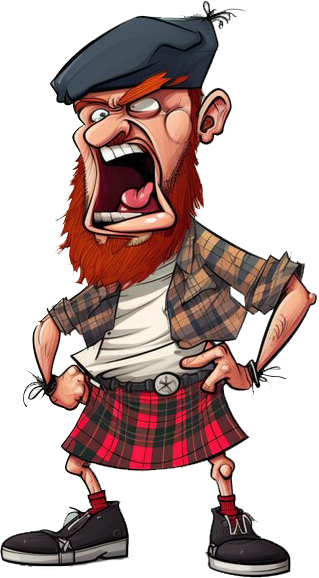}} & \textit{Epigenetics}: Being a descendant of Sir William Armstrong's angst-prone murderous cattle-thieving clan (Sec.~\ref{sec:theory3}).  & But I don't steal chairs or murder people at Costco.  \\\bottomrule
\end{tabular}
\caption{\textbf{Possible reasons I abhor chairs and Costco}, none of which are entirely satisfying (\textit{images from top: Refs}~\cite{ivoryTower,brain,sirWilliamArmstrong}).}
\label{table1}
\end{table}

\section{Discussion}

Of the three I have superficially explored (Table~\ref{table1}), the epigenetics interpretation seems the most compelling.  Still, questions remain.

My anxiety seems to be worsening with time.  Epigenetics offers an explanation for that, in terms of a snowball effect.  Identical twins grow increasingly dissimilar as they age, as their respective epigenomes evolve and lock in (or out) certain behaviors~\cite{steves2012ageing}.  That does mirror my experience with most people I've known a long time: we all seem to be becoming more like we already were.  But I'm not deep enough into the literature to understand whether epigenetics digs at \textit{why} this trend happens.  For me, life has become increasingly stable, in terms of relationships and livelihood.  So why isn't my anxiety easing rather than intensifying? 

The answer, from all three disciplines explored in this paper, is probably something like: "one's first few years is wired in deep, and forever will be the lens through which you experience subsequent years."  But perhaps it's also that the better I have it, the more I have to lose, the more anxious I am about losing it?  And/or the less time I realize I have to have what I have?

Well, setting aside the reason for the trend: how much work would it take to mitigate it? 
\subsection*{\textbf{But do I want to change?}}
Wait -- do I \textit{want} to put any work into it?  Would I like the outcome?

I'm comfortable being uncomfortable.  Maybe this ties to why my triggers do not include truly angst-worthy circumstances.  I used to volunteer in Neurosurgery at NY Presbyterian, and it was the most grounding time of my week.  My head would clear; it steeled me for the week ahead.  It's things being "fine" that creeps me out. 

And the angst is drive.  If I found it relaxing to tune out into a television screen, would I lose my edge?  Lose the itch to write, be less compelled to do science? 

And the angst is material.  If I could handle talk therapy, would I then find nothing to write about?  Would I stop seeing black humor everywhere I look?  That sounds like death.

I deal with my pathologies with averted vision.  To stare straight at them is horrifying, but juicing them is joyous.  And I don't need comedy like I need potable water and Bandaids, but without it, why stick around?  To be fine? 
\subsection*{\textbf{And need I change?}}
And is my anxiety all that bad compared to, say, yours?  Self reports can't show that.  What if we could climb into each other's heads and measure how intrinsically unsettled we are compared to each other?

Some degree of anxiety is natural.  I live in a first-world country during (relative) peacetime.  This is not representative of the human experience.  Historically, the person who was calm at rest was probably the one who got eaten by a passing bear.  To some degree, imposter "syndrome" is a misnomer.  Post-traumatic stress "disorder": misnomer.  At what point does anxiety become dysfunctional?  How do we define "dysfunction" in a culture-agnostic way~\cite{segura2015defining}? 

Well, how about I stop here.  Because science isn't going to answer my questions, nor should it.  Also because, for me, and for the moment, pretty much everything's fine.  Transiently, preciously fine.  And I don't ever want to feel fine about that.
\section{Acknowledgements}
E.A. acknowledges NSF Grants 2139004 and 2310066.

\nocite{TitlesOn}
\bibliography{bib_AF2024}

\begin{thebibliography}{43}%
\makeatletter
\providecommand \@ifxundefined [1]{%
 \@ifx{#1\undefined}
}%
\providecommand \@ifnum [1]{%
 \ifnum #1\expandafter \@firstoftwo
 \else \expandafter \@secondoftwo
 \fi
}%
\providecommand \@ifx [1]{%
 \ifx #1\expandafter \@firstoftwo
 \else \expandafter \@secondoftwo
 \fi
}%
\providecommand \natexlab [1]{#1}%
\providecommand \enquote  [1]{``#1''}%
\providecommand \bibnamefont  [1]{#1}%
\providecommand \bibfnamefont [1]{#1}%
\providecommand \citenamefont [1]{#1}%
\providecommand \href@noop [0]{\@secondoftwo}%
\providecommand \href [0]{\begingroup \@sanitize@url \@href}%
\providecommand \@href[1]{\@@startlink{#1}\@@href}%
\providecommand \@@href[1]{\endgroup#1\@@endlink}%
\providecommand \@sanitize@url [0]{\catcode `\\12\catcode `\$12\catcode
  `\&12\catcode `\#12\catcode `\^12\catcode `\_12\catcode `\%12\relax}%
\providecommand \@@startlink[1]{}%
\providecommand \@@endlink[0]{}%
\providecommand \url  [0]{\begingroup\@sanitize@url \@url }%
\providecommand \@url [1]{\endgroup\@href {#1}{\urlprefix }}%
\providecommand \urlprefix  [0]{URL }%
\providecommand \Eprint [0]{\href }%
\providecommand \doibase [0]{https://doi.org/}%
\providecommand \selectlanguage [0]{\@gobble}%
\providecommand \bibinfo  [0]{\@secondoftwo}%
\providecommand \bibfield  [0]{\@secondoftwo}%
\providecommand \translation [1]{[#1]}%
\providecommand \BibitemOpen [0]{}%
\providecommand \bibitemStop [0]{}%
\providecommand \bibitemNoStop [0]{.\EOS\space}%
\providecommand \EOS [0]{\spacefactor3000\relax}%
\providecommand \BibitemShut  [1]{\csname bibitem#1\endcsname}%
\let\auto@bib@innerbib\@empty
\bibitem [{Cos()}]{Costco}%
  \BibitemOpen
  \href@noop {} {\bibinfo {title} {\textbf{Costco Membership-only Wholesale
  Warehouse}}},\ \bibinfo {howpublished} {\url{https://www.costco.com/}},\
  \bibinfo {note} {accessed: 2024-04-01}\BibitemShut {NoStop}%
\bibitem [{\citenamefont {Wenzel}(2017)}]{wenzel2017basic}%
  \BibitemOpen
  \bibfield  {author} {\bibinfo {author} {\bibfnamefont {A.}~\bibnamefont
  {Wenzel}},\ }\bibfield  {title} {\bibinfo {title} {\textbf{Basic strategies
  of cognitive behavioral therapy}},\ }\href@noop {} {\bibfield  {journal}
  {\bibinfo  {journal} {Psychiatric Clinics}\ }\textbf {\bibinfo {volume}
  {40}},\ \bibinfo {pages} {597} (\bibinfo {year} {2017})}\BibitemShut
  {NoStop}%
\bibitem [{\citenamefont {Hofmann}\ \emph {et~al.}(2012)\citenamefont
  {Hofmann}, \citenamefont {Asnaani}, \citenamefont {Vonk}, \citenamefont
  {Sawyer},\ and\ \citenamefont {Fang}}]{hofmann2012efficacy}%
  \BibitemOpen
  \bibfield  {author} {\bibinfo {author} {\bibfnamefont {S.~G.}\ \bibnamefont
  {Hofmann}}, \bibinfo {author} {\bibfnamefont {A.}~\bibnamefont {Asnaani}},
  \bibinfo {author} {\bibfnamefont {I.~J.}\ \bibnamefont {Vonk}}, \bibinfo
  {author} {\bibfnamefont {A.~T.}\ \bibnamefont {Sawyer}},\ and\ \bibinfo
  {author} {\bibfnamefont {A.}~\bibnamefont {Fang}},\ }\bibfield  {title}
  {\bibinfo {title} {\textbf{The efficacy of cognitive behavioral therapy}: A
  review of meta-analyses},\ }\href@noop {} {\bibfield  {journal} {\bibinfo
  {journal} {Cognitive therapy and research}\ }\textbf {\bibinfo {volume}
  {36}},\ \bibinfo {pages} {427} (\bibinfo {year} {2012})}\BibitemShut
  {NoStop}%
\bibitem [{cha()}]{chair}%
  \BibitemOpen
  \href@noop {} {\bibinfo {title} {\textbf{Lowes Violet Linen Upholstered
  Tufter Powered Reclining Recliner}}},\ \bibinfo {howpublished}
  {\url{https://www.lowes.com/}},\ \bibinfo {note} {accessed:
  2024-04-01}\BibitemShut {NoStop}%
\bibitem [{din()}]{dinnerParty}%
  \BibitemOpen
  \href@noop {} {\bibinfo {title} {Stock photo of \textbf{happy people at a
  dinner party}}},\ \bibinfo {howpublished}
  {\url{https://www.alamy.com/stock-photo/couple-looking-into-camera-party.html?page=23&sortBy=relevant}},\
  \bibinfo {note} {accessed: 2024-04-01}\BibitemShut {NoStop}%
\bibitem [{dre()}]{dress}%
  \BibitemOpen
  \href@noop {} {\bibinfo {title} {\textbf{GownTown Retro Swing Dress Cocktail
  Tea Dress}}},\ \bibinfo {howpublished} {\url{http://www.amazon.com}},\
  \bibinfo {note} {accessed: 2024-04-01}\BibitemShut {NoStop}%
\bibitem [{car()}]{car}%
  \BibitemOpen
  \href@noop {} {\bibinfo {title} {\textbf{Inside of a car}}},\ \bibinfo
  {howpublished}
  {\url{https://shop.advanceautoparts.com/r/car-culture/community/all-washed-up-cleaning-a-flood-damaged-interior}},\
  \bibinfo {note} {accessed: 2024-04-01}\BibitemShut {NoStop}%
\bibitem [{pri()}]{prison}%
  \BibitemOpen
  \href@noop {} {\bibinfo {title} {\textbf{A prison cell}}},\ \bibinfo
  {howpublished} {\url{https://robertpaterson.com/prison-cell}},\ \bibinfo
  {note} {accessed: 2024-04-01}\BibitemShut {NoStop}%
\bibitem [{\citenamefont {Friedrich}\ \emph {et~al.}(2023)\citenamefont
  {Friedrich}, \citenamefont {Bareis}, \citenamefont {Bross}, \citenamefont
  {B{\"u}rger}, \citenamefont {Cort{\'e}s~Rodr{\'\i}guez}, \citenamefont
  {Effenberger}, \citenamefont {Kleinhansl}, \citenamefont {Kremer},\ and\
  \citenamefont {Schr{\"o}der}}]{friedrich2023your}%
  \BibitemOpen
  \bibfield  {author} {\bibinfo {author} {\bibfnamefont {J.}~\bibnamefont
  {Friedrich}}, \bibinfo {author} {\bibfnamefont {A.}~\bibnamefont {Bareis}},
  \bibinfo {author} {\bibfnamefont {M.}~\bibnamefont {Bross}}, \bibinfo
  {author} {\bibfnamefont {Z.}~\bibnamefont {B{\"u}rger}}, \bibinfo {author}
  {\bibfnamefont {{\'A}.}~\bibnamefont {Cort{\'e}s~Rodr{\'\i}guez}}, \bibinfo
  {author} {\bibfnamefont {N.}~\bibnamefont {Effenberger}}, \bibinfo {author}
  {\bibfnamefont {M.}~\bibnamefont {Kleinhansl}}, \bibinfo {author}
  {\bibfnamefont {F.}~\bibnamefont {Kremer}},\ and\ \bibinfo {author}
  {\bibfnamefont {C.}~\bibnamefont {Schr{\"o}der}},\ }\bibfield  {title}
  {\bibinfo {title} {\textbf{“How is your thesis going?”--Ph. D.
  students’ perspectives on mental health and stress in academia}},\
  }\href@noop {} {\bibfield  {journal} {\bibinfo  {journal} {Plos one}\
  }\textbf {\bibinfo {volume} {18}},\ \bibinfo {pages} {e0288103} (\bibinfo
  {year} {2023})}\BibitemShut {NoStop}%
\bibitem [{\citenamefont {Evans}\ \emph {et~al.}(2018)\citenamefont {Evans},
  \citenamefont {Bira}, \citenamefont {Gastelum}, \citenamefont {Weiss},\ and\
  \citenamefont {Vanderford}}]{evans2018evidence}%
  \BibitemOpen
  \bibfield  {author} {\bibinfo {author} {\bibfnamefont {T.~M.}\ \bibnamefont
  {Evans}}, \bibinfo {author} {\bibfnamefont {L.}~\bibnamefont {Bira}},
  \bibinfo {author} {\bibfnamefont {J.~B.}\ \bibnamefont {Gastelum}}, \bibinfo
  {author} {\bibfnamefont {L.~T.}\ \bibnamefont {Weiss}},\ and\ \bibinfo
  {author} {\bibfnamefont {N.~L.}\ \bibnamefont {Vanderford}},\ }\bibfield
  {title} {\bibinfo {title} {Evidence for a \textbf{mental health crisis in
  graduate education}},\ }\href@noop {} {\bibfield  {journal} {\bibinfo
  {journal} {Nature biotechnology}\ }\textbf {\bibinfo {volume} {36}},\
  \bibinfo {pages} {282} (\bibinfo {year} {2018})}\BibitemShut {NoStop}%
\bibitem [{\citenamefont {Higgins}\ and\ \citenamefont
  {Kotrlik}(2006)}]{higgins2006factors}%
  \BibitemOpen
  \bibfield  {author} {\bibinfo {author} {\bibfnamefont {C.}~\bibnamefont
  {Higgins}}\ and\ \bibinfo {author} {\bibfnamefont {J.}~\bibnamefont
  {Kotrlik}},\ }\bibfield  {title} {\bibinfo {title} {Factors associated with
  \textbf{research anxiety of university human resource education faculty}},\
  }\href@noop {} {\bibfield  {journal} {\bibinfo  {journal} {Career and
  Technical Education Research}\ }\textbf {\bibinfo {volume} {31}},\ \bibinfo
  {pages} {175} (\bibinfo {year} {2006})}\BibitemShut {NoStop}%
\bibitem [{\citenamefont {Ashrafi-Rizi}\ \emph {et~al.}(2014)\citenamefont
  {Ashrafi-Rizi}, \citenamefont {Zarmehr}, \citenamefont {Bahrami},
  \citenamefont {Ghazavi-Khorasgani}, \citenamefont {Kazempour},\ and\
  \citenamefont {Shahrzadi}}]{ashrafi2014study}%
  \BibitemOpen
  \bibfield  {author} {\bibinfo {author} {\bibfnamefont {H.}~\bibnamefont
  {Ashrafi-Rizi}}, \bibinfo {author} {\bibfnamefont {F.}~\bibnamefont
  {Zarmehr}}, \bibinfo {author} {\bibfnamefont {S.}~\bibnamefont {Bahrami}},
  \bibinfo {author} {\bibfnamefont {Z.}~\bibnamefont {Ghazavi-Khorasgani}},
  \bibinfo {author} {\bibfnamefont {Z.}~\bibnamefont {Kazempour}},\ and\
  \bibinfo {author} {\bibfnamefont {L.}~\bibnamefont {Shahrzadi}},\ }\bibfield
  {title} {\bibinfo {title} {Study on \textbf{research anxiety among faculty
  members of Isfahan University of Medical Sciences}},\ }\href@noop {}
  {\bibfield  {journal} {\bibinfo  {journal} {Materia socio-medica}\ }\textbf
  {\bibinfo {volume} {26}},\ \bibinfo {pages} {356} (\bibinfo {year}
  {2014})}\BibitemShut {NoStop}%
\bibitem [{zoo()}]{zoom}%
  \BibitemOpen
  \href@noop {} {\bibinfo {title} {\textbf{Zoom}: video telephony software}},\
  \bibinfo {howpublished} {\url{https://www.zoom.us}},\ \bibinfo {note}
  {accessed: 2024-04-01}\BibitemShut {NoStop}%
\bibitem [{mov()}]{movies}%
  \BibitemOpen
  \href@noop {} {\bibinfo {title} {Stock photo of \textbf{people at the
  movies}}},\ \bibinfo {howpublished} {\url{https://www.istockphoto.com/}},\
  \bibinfo {note} {accessed: 2024-04-01}\BibitemShut {NoStop}%
\bibitem [{the()}]{therapist}%
  \BibitemOpen
  \href@noop {} {\bibinfo {title} {Stock photo of a \textbf{judgemental person
  holding a book}}},\ \bibinfo {howpublished}
  {\url{https://www.istockphoto.com/}},\ \bibinfo {note} {accessed:
  2024-04-01}\BibitemShut {NoStop}%
\bibitem [{sub()}]{subway}%
  \BibitemOpen
  \href@noop {} {\bibinfo {title} {Stock photo of \textbf{moving subway}}},\
  \bibinfo {howpublished} {\url{https://www.alamy.com/}},\ \bibinfo {note}
  {accessed: 2024-04-01}\BibitemShut {NoStop}%
\bibitem [{\citenamefont {DeYoung}\ \emph {et~al.}(2022)\citenamefont
  {DeYoung}, \citenamefont {Beaty}, \citenamefont {Gen{\c{c}}}, \citenamefont
  {Latzman}, \citenamefont {Passamonti}, \citenamefont {Servaas}, \citenamefont
  {Shackman}, \citenamefont {Smillie}, \citenamefont {Spreng}, \citenamefont
  {Viding} \emph {et~al.}}]{deyoung2022personality}%
  \BibitemOpen
  \bibfield  {author} {\bibinfo {author} {\bibfnamefont {C.~G.}\ \bibnamefont
  {DeYoung}}, \bibinfo {author} {\bibfnamefont {R.~E.}\ \bibnamefont {Beaty}},
  \bibinfo {author} {\bibfnamefont {E.}~\bibnamefont {Gen{\c{c}}}}, \bibinfo
  {author} {\bibfnamefont {R.~D.}\ \bibnamefont {Latzman}}, \bibinfo {author}
  {\bibfnamefont {L.}~\bibnamefont {Passamonti}}, \bibinfo {author}
  {\bibfnamefont {M.~N.}\ \bibnamefont {Servaas}}, \bibinfo {author}
  {\bibfnamefont {A.~J.}\ \bibnamefont {Shackman}}, \bibinfo {author}
  {\bibfnamefont {L.~D.}\ \bibnamefont {Smillie}}, \bibinfo {author}
  {\bibfnamefont {R.~N.}\ \bibnamefont {Spreng}}, \bibinfo {author}
  {\bibfnamefont {E.}~\bibnamefont {Viding}}, \emph {et~al.},\ }\bibfield
  {title} {\bibinfo {title} {\textbf{Personality neuroscience}: An emerging
  field with bright prospects},\ }\href@noop {} {\bibfield  {journal} {\bibinfo
   {journal} {Personality science}\ }\textbf {\bibinfo {volume} {3}} (\bibinfo
  {year} {2022})}\BibitemShut {NoStop}%
\bibitem [{\citenamefont {Wiggins}(1996)}]{wiggins1996five}%
  \BibitemOpen
  \bibfield  {author} {\bibinfo {author} {\bibfnamefont {J.~S.}\ \bibnamefont
  {Wiggins}},\ }\href@noop {} {\emph {\bibinfo {title} {\textbf{The five-factor
  model of personality}: Theoretical perspectives}}}\ (\bibinfo  {publisher}
  {Guilford Press},\ \bibinfo {year} {1996})\BibitemShut {NoStop}%
\bibitem [{\citenamefont {Fox}\ and\ \citenamefont
  {Shackman}(2019)}]{fox2019central}%
  \BibitemOpen
  \bibfield  {author} {\bibinfo {author} {\bibfnamefont {A.~S.}\ \bibnamefont
  {Fox}}\ and\ \bibinfo {author} {\bibfnamefont {A.~J.}\ \bibnamefont
  {Shackman}},\ }\bibfield  {title} {\bibinfo {title} {\textbf{The central
  extended amygdala in fear and anxiety}: Closing the gap between mechanistic
  and neuroimaging research},\ }\href@noop {} {\bibfield  {journal} {\bibinfo
  {journal} {Neuroscience letters}\ }\textbf {\bibinfo {volume} {693}},\
  \bibinfo {pages} {58} (\bibinfo {year} {2019})}\BibitemShut {NoStop}%
\bibitem [{\citenamefont {Saunders}\ and\ \citenamefont
  {Inzlicht}(2020)}]{saunders2020assessing}%
  \BibitemOpen
  \bibfield  {author} {\bibinfo {author} {\bibfnamefont {B.}~\bibnamefont
  {Saunders}}\ and\ \bibinfo {author} {\bibfnamefont {M.}~\bibnamefont
  {Inzlicht}},\ }\bibfield  {title} {\bibinfo {title} {\textbf{Assessing and
  adjusting for publication bias in the relationship between anxiety and the
  error-related negativity}},\ }\href@noop {} {\bibfield  {journal} {\bibinfo
  {journal} {International Journal of Psychophysiology}\ }\textbf {\bibinfo
  {volume} {155}},\ \bibinfo {pages} {87} (\bibinfo {year} {2020})}\BibitemShut
  {NoStop}%
\bibitem [{\citenamefont {Grasby}\ \emph {et~al.}(2020)\citenamefont {Grasby},
  \citenamefont {Jahanshad}, \citenamefont {Painter}, \citenamefont
  {Colodro-Conde}, \citenamefont {Bralten}, \citenamefont {Hibar},
  \citenamefont {Lind}, \citenamefont {Pizzagalli}, \citenamefont {Ching},
  \citenamefont {McMahon} \emph {et~al.}}]{grasby2020genetic}%
  \BibitemOpen
  \bibfield  {author} {\bibinfo {author} {\bibfnamefont {K.~L.}\ \bibnamefont
  {Grasby}}, \bibinfo {author} {\bibfnamefont {N.}~\bibnamefont {Jahanshad}},
  \bibinfo {author} {\bibfnamefont {J.~N.}\ \bibnamefont {Painter}}, \bibinfo
  {author} {\bibfnamefont {L.}~\bibnamefont {Colodro-Conde}}, \bibinfo {author}
  {\bibfnamefont {J.}~\bibnamefont {Bralten}}, \bibinfo {author} {\bibfnamefont
  {D.~P.}\ \bibnamefont {Hibar}}, \bibinfo {author} {\bibfnamefont {P.~A.}\
  \bibnamefont {Lind}}, \bibinfo {author} {\bibfnamefont {F.}~\bibnamefont
  {Pizzagalli}}, \bibinfo {author} {\bibfnamefont {C.~R.}\ \bibnamefont
  {Ching}}, \bibinfo {author} {\bibfnamefont {M.~A.~B.}\ \bibnamefont
  {McMahon}}, \emph {et~al.},\ }\bibfield  {title} {\bibinfo {title}
  {\textbf{The genetic architecture of the human cerebral cortex}},\
  }\href@noop {} {\bibfield  {journal} {\bibinfo  {journal} {Science}\ }\textbf
  {\bibinfo {volume} {367}},\ \bibinfo {pages} {eaay6690} (\bibinfo {year}
  {2020})}\BibitemShut {NoStop}%
\bibitem [{\citenamefont {Ahrens}\ \emph {et~al.}(2018)\citenamefont {Ahrens},
  \citenamefont {Wu}, \citenamefont {Furlan}, \citenamefont {Hwang},
  \citenamefont {Paik}, \citenamefont {Li}, \citenamefont {Penzo},
  \citenamefont {Tollkuhn},\ and\ \citenamefont {Li}}]{ahrens2018central}%
  \BibitemOpen
  \bibfield  {author} {\bibinfo {author} {\bibfnamefont {S.}~\bibnamefont
  {Ahrens}}, \bibinfo {author} {\bibfnamefont {M.~V.}\ \bibnamefont {Wu}},
  \bibinfo {author} {\bibfnamefont {A.}~\bibnamefont {Furlan}}, \bibinfo
  {author} {\bibfnamefont {G.-R.}\ \bibnamefont {Hwang}}, \bibinfo {author}
  {\bibfnamefont {R.}~\bibnamefont {Paik}}, \bibinfo {author} {\bibfnamefont
  {H.}~\bibnamefont {Li}}, \bibinfo {author} {\bibfnamefont {M.~A.}\
  \bibnamefont {Penzo}}, \bibinfo {author} {\bibfnamefont {J.}~\bibnamefont
  {Tollkuhn}},\ and\ \bibinfo {author} {\bibfnamefont {B.}~\bibnamefont {Li}},\
  }\bibfield  {title} {\bibinfo {title} {\textbf{A central extended amygdala
  circuit that modulates anxiety}},\ }\href@noop {} {\bibfield  {journal}
  {\bibinfo  {journal} {Journal of Neuroscience}\ }\textbf {\bibinfo {volume}
  {38}},\ \bibinfo {pages} {5567} (\bibinfo {year} {2018})}\BibitemShut
  {NoStop}%
\bibitem [{\citenamefont {Carey}(2012)}]{carey2012epigenetics}%
  \BibitemOpen
  \bibfield  {author} {\bibinfo {author} {\bibfnamefont {N.}~\bibnamefont
  {Carey}},\ }\href@noop {} {\emph {\bibinfo {title} {\textbf{The epigenetics
  revolution}: How modern biology is rewriting our understanding of genetics,
  disease, and inheritance}}}\ (\bibinfo  {publisher} {Columbia University
  Press},\ \bibinfo {year} {2012})\BibitemShut {NoStop}%
\bibitem [{\citenamefont {Moore}(2017)}]{moore2017behavioral}%
  \BibitemOpen
  \bibfield  {author} {\bibinfo {author} {\bibfnamefont {D.~S.}\ \bibnamefont
  {Moore}},\ }\bibfield  {title} {\bibinfo {title} {\textbf{Behavioral
  epigenetics}},\ }\href@noop {} {\bibfield  {journal} {\bibinfo  {journal}
  {Wiley Interdisciplinary Reviews: Systems Biology and Medicine}\ }\textbf
  {\bibinfo {volume} {9}},\ \bibinfo {pages} {e1333} (\bibinfo {year}
  {2017})}\BibitemShut {NoStop}%
\bibitem [{\citenamefont {Seebacher}\ and\ \citenamefont
  {Krause}(2019)}]{seebacher2019epigenetics}%
  \BibitemOpen
  \bibfield  {author} {\bibinfo {author} {\bibfnamefont {F.}~\bibnamefont
  {Seebacher}}\ and\ \bibinfo {author} {\bibfnamefont {J.}~\bibnamefont
  {Krause}},\ }\bibfield  {title} {\bibinfo {title} {\textbf{Epigenetics of
  social behaviour}},\ }\href@noop {} {\bibfield  {journal} {\bibinfo
  {journal} {Trends in ecology \& evolution}\ }\textbf {\bibinfo {volume}
  {34}},\ \bibinfo {pages} {818} (\bibinfo {year} {2019})}\BibitemShut
  {NoStop}%
\bibitem [{\citenamefont {Feil}\ and\ \citenamefont
  {Fraga}(2012)}]{feil2012epigenetics}%
  \BibitemOpen
  \bibfield  {author} {\bibinfo {author} {\bibfnamefont {R.}~\bibnamefont
  {Feil}}\ and\ \bibinfo {author} {\bibfnamefont {M.~F.}\ \bibnamefont
  {Fraga}},\ }\bibfield  {title} {\bibinfo {title} {\textbf{Epigenetics and the
  environment}: emerging patterns and implications},\ }\href@noop {} {\bibfield
   {journal} {\bibinfo  {journal} {Nature reviews genetics}\ }\textbf {\bibinfo
  {volume} {13}},\ \bibinfo {pages} {97} (\bibinfo {year} {2012})}\BibitemShut
  {NoStop}%
\bibitem [{\citenamefont {Berkel}\ and\ \citenamefont
  {Pandey}(2017)}]{berkel2017emerging}%
  \BibitemOpen
  \bibfield  {author} {\bibinfo {author} {\bibfnamefont {T.~D.}\ \bibnamefont
  {Berkel}}\ and\ \bibinfo {author} {\bibfnamefont {S.~C.}\ \bibnamefont
  {Pandey}},\ }\bibfield  {title} {\bibinfo {title} {\textbf{Emerging role of
  epigenetic mechanisms in alcohol addiction}},\ }\href@noop {} {\bibfield
  {journal} {\bibinfo  {journal} {Alcoholism: Clinical and Experimental
  Research}\ }\textbf {\bibinfo {volume} {41}},\ \bibinfo {pages} {666}
  (\bibinfo {year} {2017})}\BibitemShut {NoStop}%
\bibitem [{\citenamefont {Krishnan}\ \emph {et~al.}(2014)\citenamefont
  {Krishnan}, \citenamefont {Sakharkar}, \citenamefont {Teppen}, \citenamefont
  {Berkel},\ and\ \citenamefont {Pandey}}]{krishnan2014epigenetic}%
  \BibitemOpen
  \bibfield  {author} {\bibinfo {author} {\bibfnamefont {H.~R.}\ \bibnamefont
  {Krishnan}}, \bibinfo {author} {\bibfnamefont {A.~J.}\ \bibnamefont
  {Sakharkar}}, \bibinfo {author} {\bibfnamefont {T.~L.}\ \bibnamefont
  {Teppen}}, \bibinfo {author} {\bibfnamefont {T.~D.}\ \bibnamefont {Berkel}},\
  and\ \bibinfo {author} {\bibfnamefont {S.~C.}\ \bibnamefont {Pandey}},\
  }\bibfield  {title} {\bibinfo {title} {\textbf{The epigenetic landscape of
  alcoholism}},\ }\href@noop {} {\bibfield  {journal} {\bibinfo  {journal}
  {International Review of Neurobiology}\ }\textbf {\bibinfo {volume} {115}},\
  \bibinfo {pages} {75} (\bibinfo {year} {2014})}\BibitemShut {NoStop}%
\bibitem [{\citenamefont {Weaver}\ \emph {et~al.}(2004)\citenamefont {Weaver},
  \citenamefont {Cervoni}, \citenamefont {Champagne}, \citenamefont
  {D'Alessio}, \citenamefont {Sharma}, \citenamefont {Seckl}, \citenamefont
  {Dymov}, \citenamefont {Szyf},\ and\ \citenamefont
  {Meaney}}]{weaver2004epigenetic}%
  \BibitemOpen
  \bibfield  {author} {\bibinfo {author} {\bibfnamefont {I.~C.}\ \bibnamefont
  {Weaver}}, \bibinfo {author} {\bibfnamefont {N.}~\bibnamefont {Cervoni}},
  \bibinfo {author} {\bibfnamefont {F.~A.}\ \bibnamefont {Champagne}}, \bibinfo
  {author} {\bibfnamefont {A.~C.}\ \bibnamefont {D'Alessio}}, \bibinfo {author}
  {\bibfnamefont {S.}~\bibnamefont {Sharma}}, \bibinfo {author} {\bibfnamefont
  {J.~R.}\ \bibnamefont {Seckl}}, \bibinfo {author} {\bibfnamefont
  {S.}~\bibnamefont {Dymov}}, \bibinfo {author} {\bibfnamefont
  {M.}~\bibnamefont {Szyf}},\ and\ \bibinfo {author} {\bibfnamefont {M.~J.}\
  \bibnamefont {Meaney}},\ }\bibfield  {title} {\bibinfo {title}
  {\textbf{Epigenetic programming by maternal behavior}},\ }\href@noop {}
  {\bibfield  {journal} {\bibinfo  {journal} {Nature neuroscience}\ }\textbf
  {\bibinfo {volume} {7}},\ \bibinfo {pages} {847} (\bibinfo {year}
  {2004})}\BibitemShut {NoStop}%
\bibitem [{\citenamefont {Puglia}\ \emph {et~al.}(2015)\citenamefont {Puglia},
  \citenamefont {Lillard}, \citenamefont {Morris},\ and\ \citenamefont
  {Connelly}}]{puglia2015epigenetic}%
  \BibitemOpen
  \bibfield  {author} {\bibinfo {author} {\bibfnamefont {M.~H.}\ \bibnamefont
  {Puglia}}, \bibinfo {author} {\bibfnamefont {T.~S.}\ \bibnamefont {Lillard}},
  \bibinfo {author} {\bibfnamefont {J.~P.}\ \bibnamefont {Morris}},\ and\
  \bibinfo {author} {\bibfnamefont {J.~J.}\ \bibnamefont {Connelly}},\
  }\bibfield  {title} {\bibinfo {title} {\textbf{Epigenetic modification of the
  oxytocin receptor gene influences the perception of anger and fear in the
  human brain}},\ }\href@noop {} {\bibfield  {journal} {\bibinfo  {journal}
  {Proceedings of the National Academy of Sciences}\ }\textbf {\bibinfo
  {volume} {112}},\ \bibinfo {pages} {3308} (\bibinfo {year}
  {2015})}\BibitemShut {NoStop}%
\bibitem [{\citenamefont {Dias}\ and\ \citenamefont
  {Ressler}(2014)}]{dias2014parental}%
  \BibitemOpen
  \bibfield  {author} {\bibinfo {author} {\bibfnamefont {B.~G.}\ \bibnamefont
  {Dias}}\ and\ \bibinfo {author} {\bibfnamefont {K.~J.}\ \bibnamefont
  {Ressler}},\ }\bibfield  {title} {\bibinfo {title} {\textbf{Parental
  olfactory experience influences behavior and neural structure in subsequent
  generations}},\ }\href@noop {} {\bibfield  {journal} {\bibinfo  {journal}
  {Nature neuroscience}\ }\textbf {\bibinfo {volume} {17}},\ \bibinfo {pages}
  {89} (\bibinfo {year} {2014})}\BibitemShut {NoStop}%
\bibitem [{\citenamefont {Krippner}\ and\ \citenamefont
  {Barrett}(2019)}]{krippner2019transgenerational}%
  \BibitemOpen
  \bibfield  {author} {\bibinfo {author} {\bibfnamefont {S.}~\bibnamefont
  {Krippner}}\ and\ \bibinfo {author} {\bibfnamefont {D.}~\bibnamefont
  {Barrett}},\ }\bibfield  {title} {\bibinfo {title} {\textbf{Transgenerational
  Trauma}},\ }\href@noop {} {\bibfield  {journal} {\bibinfo  {journal} {The
  Journal of Mind and Behavior}\ }\textbf {\bibinfo {volume} {40}},\ \bibinfo
  {pages} {53} (\bibinfo {year} {2019})}\BibitemShut {NoStop}%
\bibitem [{\citenamefont {Rutledge}(1966)}]{rutledge1966lawless}%
  \BibitemOpen
  \bibfield  {author} {\bibinfo {author} {\bibfnamefont {J.~E.}\ \bibnamefont
  {Rutledge}},\ }\bibfield  {title} {\bibinfo {title} {{\textbf{The Lawless
  Clan: The Armstrongs}}},\ }\href@noop {} {\bibfield  {journal} {\bibinfo
  {journal} {The Dalhousie Review}\ } (\bibinfo {year} {1966})}\BibitemShut
  {NoStop}%
\bibitem [{\citenamefont {Wright}(2022)}]{wright2022nature}%
  \BibitemOpen
  \bibfield  {author} {\bibinfo {author} {\bibfnamefont {R.~O.}\ \bibnamefont
  {Wright}},\ }\bibfield  {title} {\bibinfo {title} {\textbf{Nature versus
  nurture—on the origins of a specious argument}},\ }\href@noop {} {\bibfield
   {journal} {\bibinfo  {journal} {Exposome}\ }\textbf {\bibinfo {volume}
  {2}},\ \bibinfo {pages} {osac005} (\bibinfo {year} {2022})}\BibitemShut
  {NoStop}%
\bibitem [{ang()}]{anglican}%
  \BibitemOpen
  \href@noop {} {\bibinfo {title} {{\textbf{Anglican Pacifist Fellowship}}}},\
  \bibinfo {howpublished} {\url{https://www.anglicanpeacemaker.org.uk/}},\
  \bibinfo {note} {accessed: 2024-04-01}\BibitemShut {NoStop}%
\bibitem [{\citenamefont {Stewart}(2017)}]{stewart2017armstrongs}%
  \BibitemOpen
  \bibfield  {author} {\bibinfo {author} {\bibfnamefont {D.~J.}\ \bibnamefont
  {Stewart}},\ }\href@noop {} {\emph {\bibinfo {title} {\textbf{The
  Armstrongs}}}}\ (\bibinfo  {publisher} {American Academic Press},\ \bibinfo
  {year} {2017})\BibitemShut {NoStop}%
\bibitem [{\citenamefont {Mackenzie}(1879)}]{mackenzie1879history}%
  \BibitemOpen
  \bibfield  {author} {\bibinfo {author} {\bibfnamefont {A.}~\bibnamefont
  {Mackenzie}},\ }\href@noop {} {\emph {\bibinfo {title} {\textbf{History of
  the Clan Mackenzie}. With Genealogies of the Principal Families of the
  Name}}}\ (\bibinfo  {publisher} {A. \& W. Mackenzie},\ \bibinfo {year}
  {1879})\BibitemShut {NoStop}%
\bibitem [{coa()}]{coatOfArms}%
  \BibitemOpen
  \href@noop {} {\bibinfo {title} {\textbf{Armstrong coat of arms}}},\ \bibinfo
  {howpublished} {\url{https://scotstee.com/}},\ \bibinfo {note} {accessed:
  2024-04-01}\BibitemShut {NoStop}%
\bibitem [{ivo()}]{ivoryTower}%
  \BibitemOpen
  \href@noop {} {\bibinfo {title} {\textbf{A tower that appears ivory}, at an
  undisclosed location}},\ \bibinfo {howpublished}
  {\url{https://www.cam.ac.uk/research/news/exploding-the-ivory-tower-myth}},\
  \bibinfo {note} {accessed: 2024-04-01}\BibitemShut {NoStop}%
\bibitem [{bra()}]{brain}%
  \BibitemOpen
  \href@noop {} {\bibinfo {title} {Location of the \textbf{human dorsomedial
  prefrontal cortex}}},\ \bibinfo {howpublished}
  {\url{https://en.wikipedia.org/wiki/Dorsomedial_prefrontal_cortex}},\
  \bibinfo {note} {accessed: 2024-04-01}\BibitemShut {NoStop}%
\bibitem [{sir()}]{sirWilliamArmstrong}%
  \BibitemOpen
  \href@noop {} {\bibinfo {title} {Stock cartoon of a \textbf{Scottish
  gangster}}},\ \bibinfo {howpublished} {\url{https://www.freepik.com/}},\
  \bibinfo {note} {accessed: 2024-04-01}\BibitemShut {NoStop}%
\bibitem [{\citenamefont {Steves}\ \emph {et~al.}(2012)\citenamefont {Steves},
  \citenamefont {Spector},\ and\ \citenamefont {Jackson}}]{steves2012ageing}%
  \BibitemOpen
  \bibfield  {author} {\bibinfo {author} {\bibfnamefont {C.~J.}\ \bibnamefont
  {Steves}}, \bibinfo {author} {\bibfnamefont {T.~D.}\ \bibnamefont
  {Spector}},\ and\ \bibinfo {author} {\bibfnamefont {S.~H.}\ \bibnamefont
  {Jackson}},\ }\bibfield  {title} {\bibinfo {title} {\textbf{Ageing, genes,
  environment and epigenetics: what twin studies tell us} now, and in the
  future},\ }\href@noop {} {\bibfield  {journal} {\bibinfo  {journal} {Age and
  ageing}\ }\textbf {\bibinfo {volume} {41}},\ \bibinfo {pages} {581} (\bibinfo
  {year} {2012})}\BibitemShut {NoStop}%
\bibitem [{\citenamefont {Segura}(2015)}]{segura2015defining}%
  \BibitemOpen
  \bibfield  {author} {\bibinfo {author} {\bibfnamefont {K.}~\bibnamefont
  {Segura}},\ }\href@noop {} {\bibinfo {title} {\textbf{Defining normal}}}
  (\bibinfo {year} {2015})\BibitemShut {NoStop}%
\end{thebibliography}%


%


\end{document}